\documentclass[twocolumn,titlepage,preprintnumbers,amsmath,amssymb,superscriptaddress,10pt]{revtex4-1}
\usepackage[dvips]{graphicx}
\usepackage{bm}
\usepackage{amssymb}
\usepackage{latexsym}
\usepackage{hyperref}
\usepackage{ulem}
\usepackage{color}
\usepackage{amsmath}
\usepackage{lipsum}
\usepackage{floatrow}

\begin{document}
\title{Self-amplification of solid friction in interleaved assemblies}
\author{H\'ector Alarc\'on}
\affiliation{Laboratoire de Physique des Solides, CNRS \& Universit\'e Paris-Sud, Universit\'e Paris-Saclay, 91405 Orsay Cedex, France}
\affiliation{Laboratoire de Physico-Chimie Th\'eorique, UMR CNRS 7083 Gulliver, ESPCI ParisTech, PSL Research University, 75005 Paris, France}
\author{Thomas Salez}
\affiliation{Laboratoire de Physico-Chimie Th\'eorique, UMR CNRS 7083 Gulliver, ESPCI ParisTech, PSL Research University, 75005 Paris, France}
\author{Christophe Poulard}
\affiliation{Laboratoire de Physique des Solides, CNRS \& Universit\'e Paris-Sud, Universit\'e Paris-Saclay, 91405 Orsay Cedex, France}
\author{Jean-Francis Bloch}
\affiliation{Universit\'e de Grenoble Alpes, LGP2, CS 10065, 38402 Saint-Martin-d'H\`eres, France}
\author{\'Elie Rapha\"{e}l}
\affiliation{Laboratoire de Physico-Chimie Th\'eorique, UMR CNRS 7083 Gulliver, ESPCI ParisTech, PSL Research University, 75005 Paris, France}
\author{Kari Dalnoki-Veress}
\affiliation{Laboratoire de Physico-Chimie Th\'eorique, UMR CNRS 7083 Gulliver, ESPCI ParisTech, PSL Research University, 75005 Paris, France}
\affiliation{Department of Physics $\&$ Astronomy, McMaster University, Hamilton, ON, Canada L8S 4M1}
\author{Fr\'ed\'eric Restagno}\email{frederic.restagno@u-psud.fr}
\affiliation{Laboratoire de Physique des Solides, CNRS \& Universit\'e Paris-Sud, Universit\'e Paris-Saclay, 91405 Orsay Cedex, France}

\begin{abstract}
It is nearly impossible to separate two interleaved phonebooks when held by their spines. A full understanding of this astonishing demonstration of solid friction in complex assemblies has remained elusive. In this Letter, we report on experiments with controlled booklets and show that the force required increases sharply with the number of sheets. A model captures the effect of the number of sheets, their thickness and the overlapping distance. Furthermore, the data collapse onto a self-similar master curve with one dimensionless amplification parameter. In addition to solving a long-standing familiar enigma, this model system provides a framework with which one can accurately measure friction forces and coefficients at low loads, and that has relevance to complex assemblies from the macro to the nanoscale. 
\end{abstract}
\maketitle

Many of us are familiar with a classical demonstration of the strength of friction: take two phonebooks, interleave their sheets and try to separate them by pulling on their spines. This demonstration has been carried out spectacularly by attempting to pull the books apart with people, trucks, lifting a car~\cite{cobayes_phonebook}, and even two military tanks~\cite{discov_phonebook}, only to fail and suggest that the inner friction between these sheets prevails. The simple explanation often given is that gravity provides the normal force that generates the tangential friction, but this hypothesis is easily proven to be wrong as there is no discernible difference between such an experiment carried out in the vertical or horizontal direction. In this Letter, we study the force needed to separate two books as a function of the number of sheets, the thickness of the sheets, and the interleaving distance. In particular, we show that the force required to separate the books increases abruptly with the number of sheets. The strength of the system is due to the operator: the person, car, truck, or tank, amplifies any small friction arising from the normal force acting on the boundaries of the stack. We present a simple model that captures all the data into a self-similar master curve. The model  depends on one single dimensionless amplification parameter, and thus gives insight into the mechanisms at play in this deceivingly complex system. In addition to solving a long-standing familiar enigma related to the classical problem of friction, this model system provides a framework within which one can accurately measure friction forces and coefficients at low loads, and opens the way to the technologically-relevant engineering of friction in complex assemblies from the macro to the nanoscale.

The first-known systematic studies of friction were carried  out five centuries ago by da Vinci~\cite{dowson1979history,krim1996friction} who discovered basic rules which were later confirmed by Amontons~\cite{dowson1979history}. In particular, these laws establish that the friction force is independent of  the contact area and proportional to the applied load during sliding, the proportionality constant being the coefficient of kinetic friction. Coulomb rediscovered these laws and further determined that, during sliding, friction is independent of the relative speed between the surfaces~\cite{dowson1979history}. This simple set of rules, collectively known as Amontons-Coulomb (AC) laws, has been well studied in macroscopic experiments over the centuries. During the last decades, efforts on the micro and nanoscale, and towards biology~\cite{ward_solid_2015}, have resulted in a resurgence of activity in tribology. For example, tools like the surface force apparatus and the atomic force microscope have fuelled experimental efforts~\cite{bhushan1995nanotribology, urbakh2004nonlinear, reiter1994static}, as well as advanced theoretical treatments that go well beyond phenomenology~\cite{muser2003statistical, muser2002nature, vanossi2011modeling}. Interest in the development of micro-electro-mechanical-systems and mechanical devices that operate on small length scales has driven much of this research. At the extreme limit, down to the nanoscale, it was found that the energy dissipation in friction depends on both electronic and phononic contributions, and that differences in the electron-phonon coupling between single and bilayer sheets of graphene result in variations in friction~\cite{filleter2009friction}. Furthermore, friction was probed in experiments on multi-walled carbon nanotubes and boron nitride nanotubes~\cite{cumings2000low, kis2006interlayer, nigues2014ultrahigh}. In such investigations, the inner tubes could be slid out of the outer tube revealing vanishingly-small molecular friction for carbon, and a much stronger, area-dependent, molecular friction for boron nitride~\cite{nigues2014ultrahigh}. Clearly, such works reveal drastic departures from the simple AC laws, as one approaches the nanoscale. Moreover, friction can be further complicated by a stick-slip response, as seen in many cases~\cite{leine1998stick, baumberger2006solid}, including layers of paper~\cite{garoff2002friction, heslot1994creep}. Remarkably, although these complex non-linear phenomena may be observed at micrometric scales, it is often the case on larger length scales that the simple AC laws with which we are so familiar are valid (see reviews~\cite{gao2004, baumberger2006solid, urbakh2004nonlinear, muser2003statistical,fulleringer2015forced}).

\begin{figure}[t!]
\begin{center}
\includegraphics[width=1\textwidth]{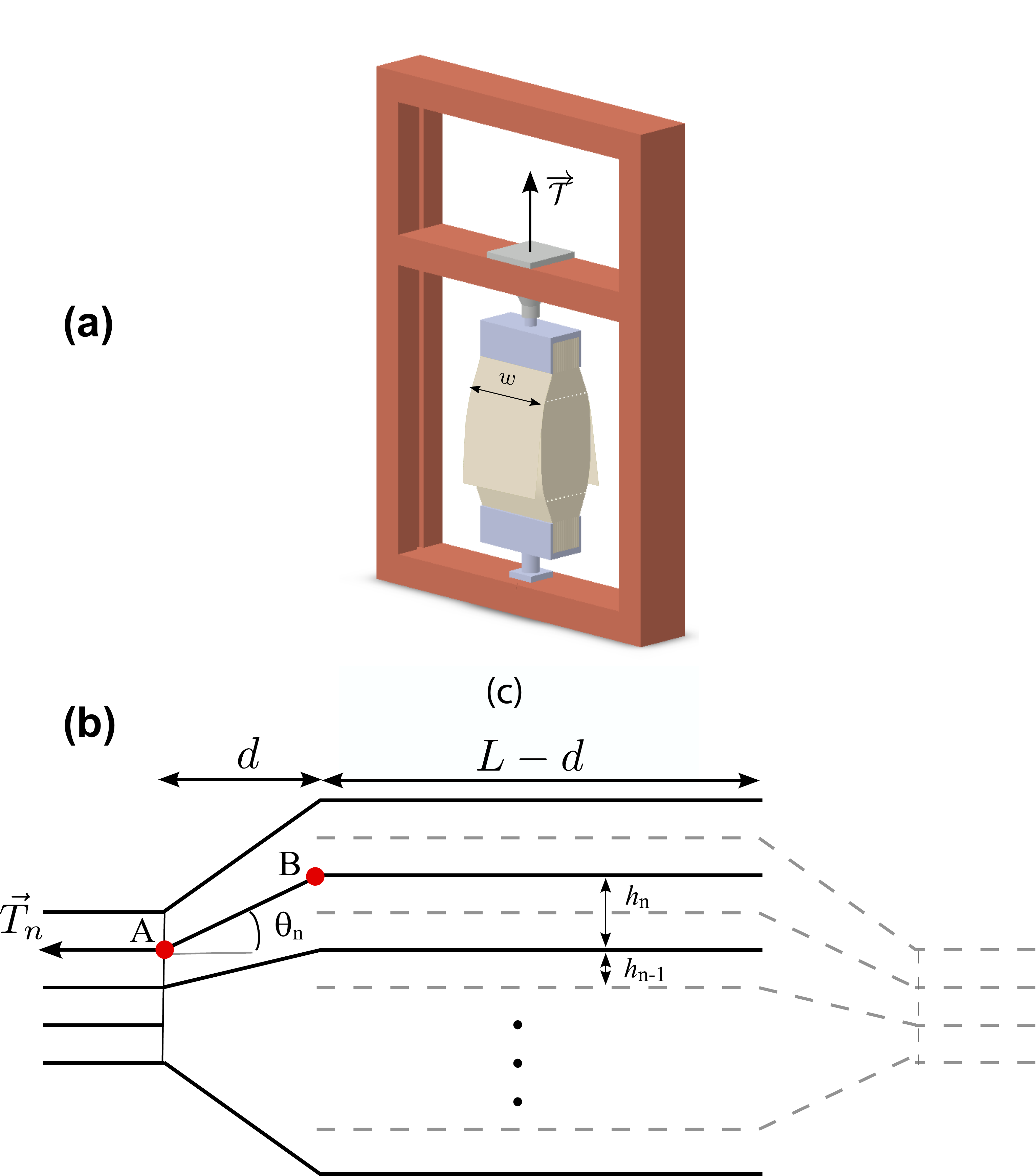}
\caption{\textit{The enigma of the two interleaved phonebooks: experimental setup, and theoretical model. a) Schematic of the experimental setup. b) Schematic of the interleaved books introducing the local traction force  $T_n$ exerted by the operator on the $n^{\textrm{th}}$ paper sheet. The two vertical lines represent the clamps, while the other lines represent the central lines of each sheet.}}
\label{fig1}
\end{center}
\end{figure}

In this Letter, we focus on the stubborn popular enigma of the two interleaved phonebooks~\cite{cobayes_phonebook,discov_phonebook}. In order to investigate these striking observations, we developed a well-controlled system of two identical books, both made up of $2M\gg1$ identical sheets of paper (Inacopia\textsuperscript{TM}) with width $w=12$~cm, length $L=25$~cm, and thickness $\epsilon=0.10$~mm. The two books were prepared by stacking sheets and holding the assembly at one end with a rigid aluminium clamp. They were perfectly interleaved sheet-by-sheet, and mounted by their clamps vertically in a traction instrument (Adamel Lhomargy DY32). Then, they were separated in a vertical orientation, while the total traction force $\mathcal{T}$ was measured (see Fig.~\ref{fig1}(a)). Since the experiments explored different ranges of force, three sensors were used with maximum forces of $10$~N, $100$~N, and $1000$~N. The length of overlap between the two books is denoted by $L-d$, where $d$ is the separation distance (measured with an accuracy of 10~$\mu$m) from the clamp of each book to the contact zone (see Fig.~\ref{fig1}(b)). Consistent with the actual experimental parameters, we make the simplifying assumption that $d$ is large compared to the total thickness $2M\epsilon$ of one book, so that the angle $\theta_n$ made by the $n^{\textrm{th}}$ sheet as it traverses from the clamp to the contact zone is small. Therefore, $L-d$ is nearly identical for all sheets. Finally, the books are separated at constant velocity (typically $1$~mm/min) and we have found, in accordance with the AC laws, that the velocity does not significantly affect the results.

\begin{figure}[t!]
 \includegraphics[width=1\textwidth]{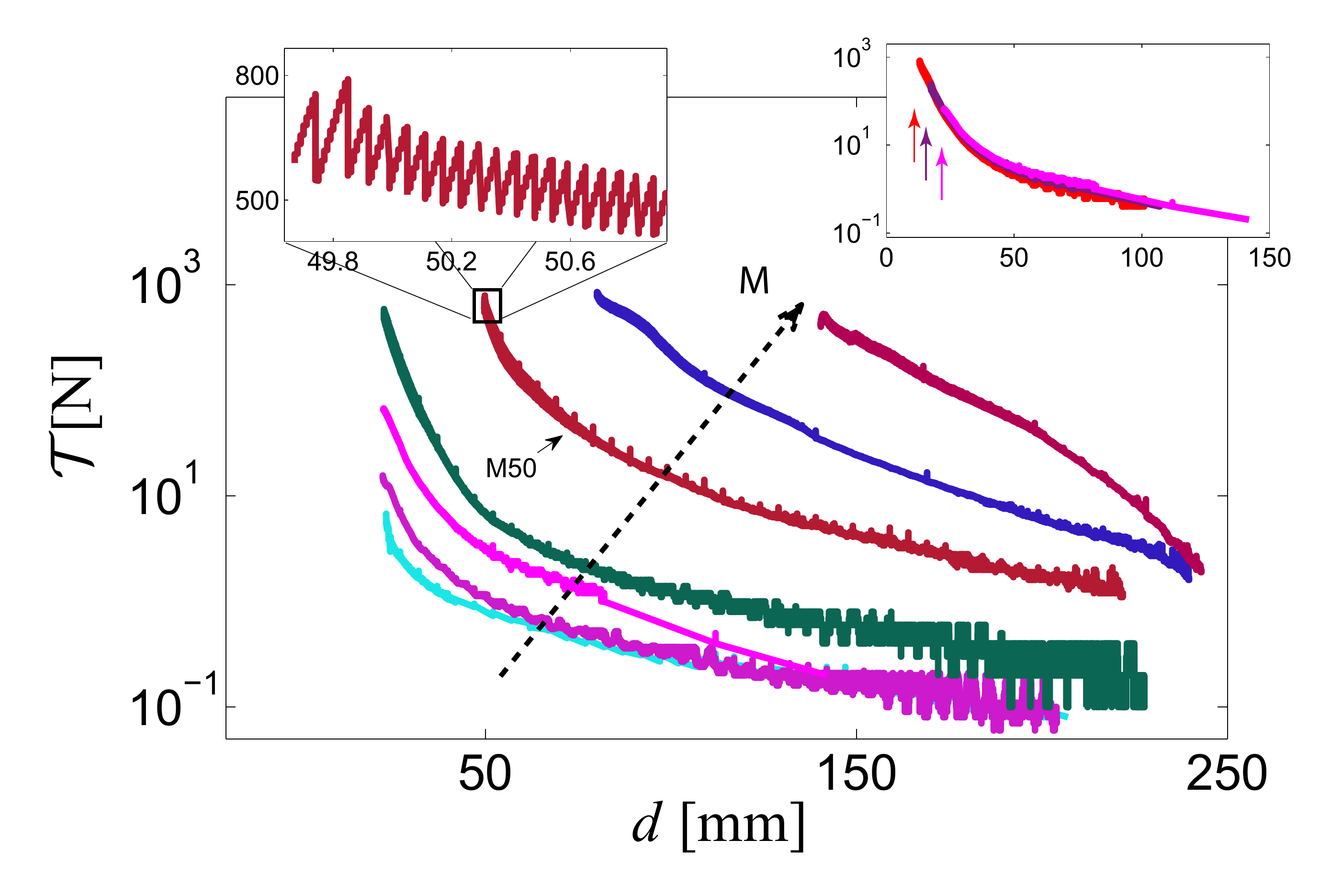}
\caption{\textit{Measured total traction force as a function of separation distance for various numbers of sheets. The number of sheets for each of the two identical books is $2M$, where $M=12,15, 23, 27, 50, 75$, and  100, with $M$ increasing as indicated by the dashed arrow (for M=50 and 100 only one of two data sets are displayed for clarity). The left inset shows a zoomed region for $M=50$ highlighting stick-slip friction. The right inset shows three experiments with $M=23$, and varying initial separation distances $d_0=11, 17$ and, 22~mm (vertical arrows).}}
\label{fig2}
\end{figure}

As the books are pulled apart, an initial maximum traction force is first reached before they start to slide with respect to each other. In Fig.~\ref{fig2}, we show the raw total traction force $\mathcal{T}$, measured during constant-velocity sliding, as a function of the distance $d$, for seven experiments with $M$ ranging from 12 to 100. As observed, the smaller $d$ or the larger $M$, the larger the traction, and those dependences are highly nonlinear. Furthermore, the amplification of friction is far from being a small effect: a single experiment spans over three decades in the traction force. Additionally, a tenfold increase in the number of sheets (\textit{e.g.} $M=12$ and $M=100$) induces a four orders of magnitude increase in the traction force. In the left inset of Fig.~\ref{fig2}, we can see clear evidence of stick-slip~\cite{garoff2002friction, heslot1994creep} for an experiment with $M=50$. However, the difference between the local maxima and minima, which results from the difference between the coefficients of static and kinetic friction, is negligible in comparison to the global amplitude in $\mathcal{T}$. We can thus neglect this effect, as well as  the difference in the coefficients of static and kinetic friction. Furthermore, in the right inset of  Fig.~\ref{fig2}, we see that the experiments are reproducible and independent of the initial separation distance $d_0$. Finally, the friction of paper can be affected by humidity changes~\cite{crassous1999humidity}. All the experiments were carried at ambient humidity which can vary from 30 to 60 \% but is usually closer to 45 \%. The day-to-day variations are negligible as can be seen from the reproducibility of the experiments.

These results can be explained using simple geometrical and mechanical arguments, for which Fig.~\ref{fig1}(b) provides detailed notations. Each sheet of a given book is indexed by $n$ ranging from $n=1$ in the middle of the book to $n=M$ at one extremity. The problem is symmetric with respect to the central line of the book. We define $H_n\equiv h_n/d=n\epsilon/d$, where $h_n$ is the shift in position of the $n^{\textrm{th}}$ sheet in the contact zone with respect to its position of clamping. As a consequence, the tilt angle $\theta_n$ of the $n^{\textrm{th}}$ sheet satisfies $\sin\theta_n=H_n/\sqrt{1+H_n^2}$ and $\cos\theta_n=1/\sqrt{1+H_n^2}$. Essentially, the tilting of each individual sheet $n$ in the intermediate region between the clamping and contact zones converts part of the local traction $T_n$ exerted on it by the operator (at point $A$) into a supplementary local normal force $T_n\tan\theta_n=H_nT_n$ exerted on the stack below it (at point $B$). Therefore, according to AC laws, this leads to a self-induced additional  inner friction force that resists the traction: the more the operator pulls, the higher the frictional resistance. At onset of sliding, the change in traction with $n$ thus reads:
\begin{equation}
T_{n}-T_{n+1}=4\mu H_n \,T_n\ ,
\label{eq1}
\end{equation}
where $\mu$ is the coefficient of kinetic friction, and the two factors of 2 come from the identical contributions of the two books and the two pages of one sheet. Finally, the boundary condition is given by $T_M=T^*$, where $T^*$ is the unknown traction exerted on the outer sheet, and the total traction force is defined by $\mathcal{T}=2\sum_{k=1}^MT_k$.

We now introduce the variable $z=n/M$, as well as the dimensionless amplification parameter $\alpha=2\mu\epsilon M^2/d$, and use a continuous description by replacing $T_n$ by $T(z)$, since $M\gg1$. We thus obtain the ODE:
\begin{equation}
T'(z)+2\alpha z\ T(z)=0\ .
\label{eq2}
\end{equation}
Together with the boundary condition $T(1)=T^*$, we get by integration the local traction force $T(z)=T^*\exp[\alpha(1-z^2)]$. Finally, we obtain the total traction force $\mathcal{T}=2M\int_0^1dz\,T(z)$ in the self-similar form:
\begin{equation}
\label{final}
\frac{\mathcal{T}}{2MT^*}=\sqrt{\frac{\pi}{4\alpha}}\ \exp(\alpha)\  \textrm{erf}\left(\sqrt{\alpha}\right)\ .
\end{equation}
This expression is remarkable for two reasons. First, it tends to 1 as $\alpha\rightarrow0$, which means that it represents the geometrical amplification gain of friction with respect to a simple linear collection of $2M$ independent flat sheets with local friction $T^*$. Secondly, it depends solely and almost exponentially on the amplification parameter $\alpha$, which thus appears as the central dimensionless parameter of this study.

\begin{figure}[t!]
\includegraphics[width=1\textwidth]{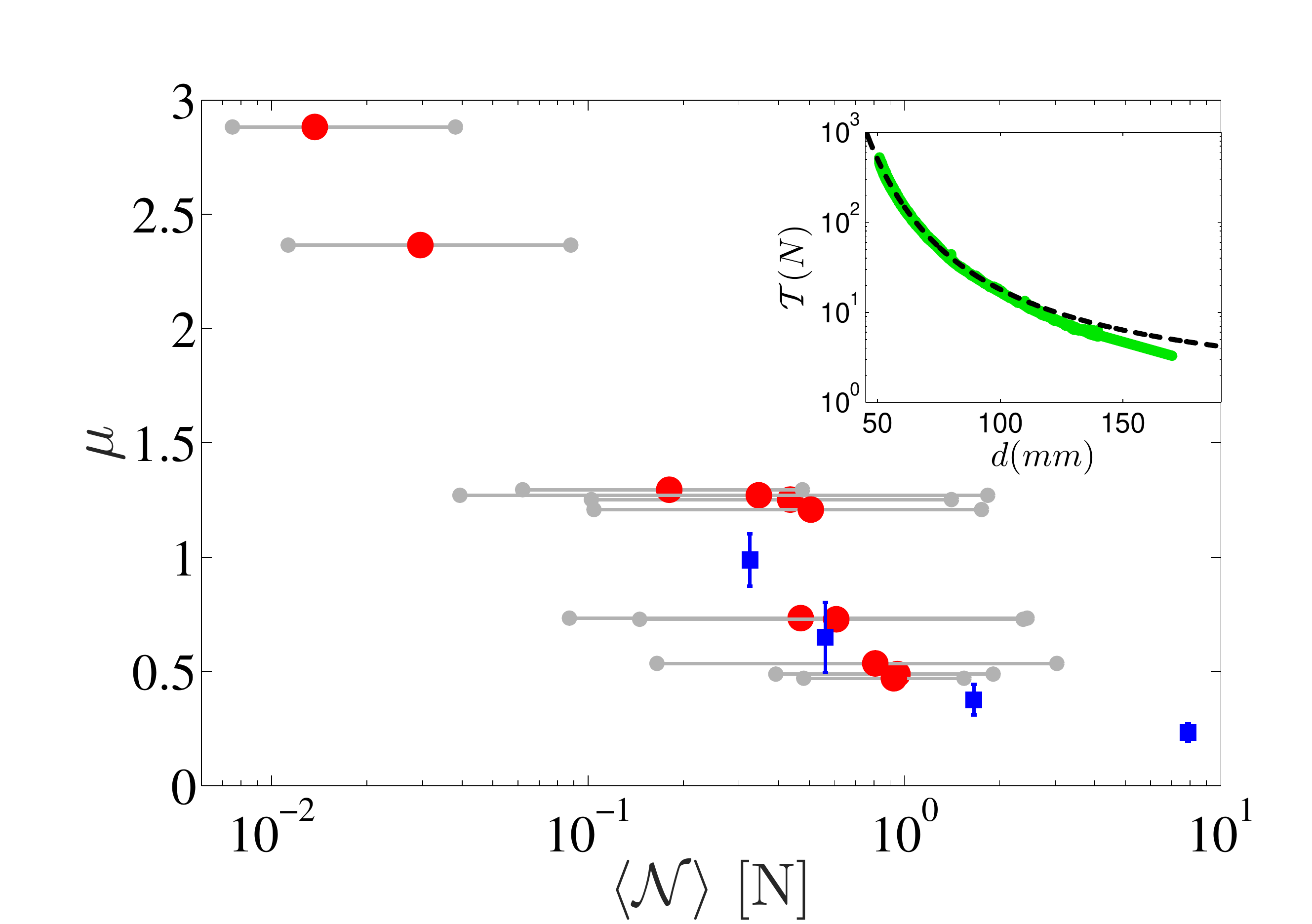}
\caption{\textit{Coefficient of kinetic friction as a function of the average effective load per page. (circles) Values of $\mu$ obtained by fitting each experiment performed on the interleaved books to Eq.~(\ref{final}), as a function of the average effective load $\langle\mathcal{N}\rangle$ defined in text. Since $\mu$ is load-dependent, we indicate the traction range over which Eq.~(\ref{final}) was fit to the data by the grey horizontal lines. (squares) Values of $\mu$ obtained independently using a tribometer for two sheets of the same paper. In this case, $\langle\mathcal{N}\rangle$ is simply the externally-applied load, and the error bars correspond to the standard deviation of the measurements that have been repeated at least five times. The inset shows a fit to~Eq.~(\ref{final}) (dashed line) of a traction experiment (plain line), for $M=50$. }}
\label{fig3}
\end{figure}

We find that the data is well described by Eq.~(\ref{final}), using a single value $T^*=0.01$~N of the microscopic traction force, and with a single coefficient of kinetic friction fit freely to each traction experiment. We note that $\mu$ is not strictly constant during an experiment since the load varies, as explained below. Taking $\mu$ to be constant is a simplifying assumption in the model and it captures the average value of $\mu$ over the range in load. A typical example is shown in the inset of Fig.~\ref{fig3}, for which an excellent fit is obtained over more than two decades in the traction force, with a constant value $\mu =0.73 \pm 0.02$. The range over which our simple model is applicable, with a constant $\mu$, is clear from the fit, and deviations are observed at small overlap between the books. Note that the tiny force $T^*$ acting on the outmost sheets is a crucial boundary condition. It is this finite force, corresponding to no more than the weight of a butterfly, that is self-amplified by the operator -- either in a well-controlled experiment as we have carried out here, or when lifting a car with phonebooks~\cite{cobayes_phonebook}. In our experiment, the boundary force $T^*$ originates from the elasticity of the paper: the outer sheets have a tendency to be flat and resist slightly the bowing induced in the contact region, thus creating a small normal force resulting in a small friction force.

The coefficient of kinetic friction is a phenomenological quantity whose value depends on several parameters~\cite{ben2011static}. For metal-on-metal~\cite{etsion1993effect} and paper-on-paper~\cite{bruker}, $\mu$ is known to be dependent on the normal load. Specifically, $\mu$ increases with decreasing load and can even be higher than 1 for small loads, due to adhesion forces~\cite{crassous1999humidity}. This fact is confirmed in Fig.~\ref{fig3}, where we plot the coefficient of kinetic friction obtained from a standard tribology experiment between two sheets of paper, as a function of the applied normal load (squares). These values were obtained at 23 $^\text{o}$C and with a relative humidity of 50 \%. Let us now compare this result with the interleaved-book case. As explained above, our model is based on the AC laws and considers a single average $\mu$ for each experiment, that is constant irrespective of the load. This hypothesis is equivalent to taking a coarse-grained approach, even though the load increases towards the centre of the books and at small $d$. The previous assumption is sufficient to describe the data, as evidenced by the fit in the inset of Fig.~\ref{fig3}. Then, we define through AC laws an effective load per page, $\mathcal{N}=\mathcal{T}/(4M\mu)$, which would correspond to the load applied on each page if the $2M$ sheets were parallel (\textit{i.e.} all $\theta_n=0$). Note that, by defining the microscopic load $\mathcal{N}^*\equiv T^*/(2\mu)$, Eq.~(\ref{final}) also provides the amplification gain $\mathcal{N}/\mathcal{N}^*$ in normal load. The average traction force $\langle \mathcal{T} \rangle$ of a given experiment is calculated over the range of applicability of the theory, and used to obtain the average effective load per page $\langle \mathcal{N} \rangle$. The result $\mu(\langle \mathcal{N} \rangle)$ is shown in Fig.~\ref{fig3} (circles), with the ranges of applicability indicated by the grey horizontal lines. In short, a single value of $\mu$ could be used to adequately describe each $\mathcal{T}(d)$ experiment with the theory, even though the load varied  over the indicated range. The consistency between the simple tribology data and that of the interleaved books provides confidence in the model. Furthermore, it is noteworthy that the classic phonebook demonstration probes a broad range of loads, that typically consists of two decades, and a nearly sixfold change in the coefficient of kinetic friction.

\begin{figure}[t!]
\includegraphics[width=1\textwidth]{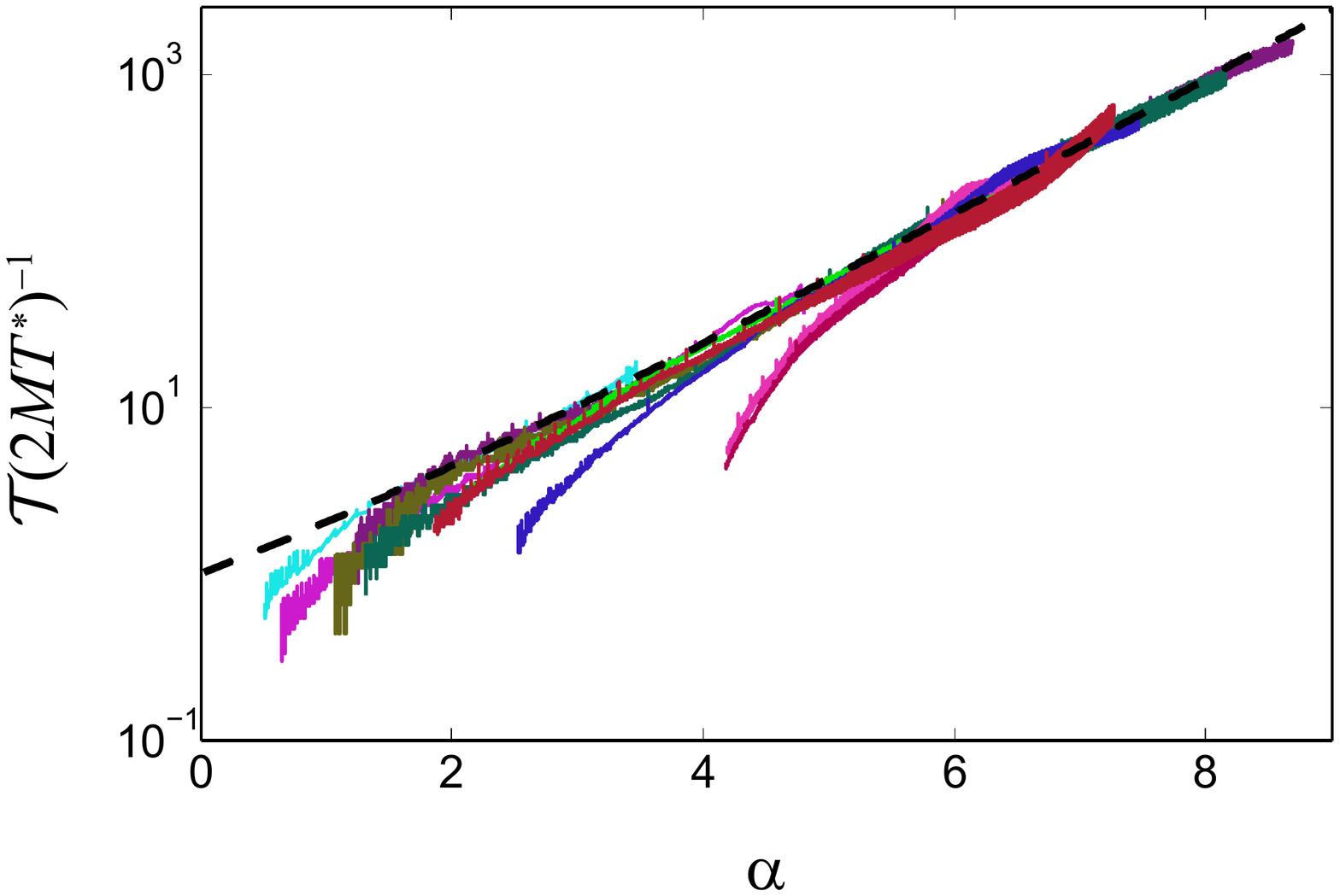}
\caption{\textit{Self-similar master curve of the rescaled total traction force as a function of the amplification parameter. To produce this figure, each of the data sets shown in Fig.~\ref{fig2} (same colour code) was fitted to Eq.~(\ref{final}) (dashed line), using  $T^*=0.01$~N, and $\mu$ as a free parameter (see Fig.~\ref{fig3}).}}
\label{fig4}
\end{figure}

The self-similar Eq.~(\ref{final}) suggests that with $T^*=0.01$~N, and the best fit values of $\mu$ for each experiment (see Fig.~\ref{fig3}), all data should collapse when normalised appropriately. Indeed, in Fig.~\ref{fig4}, we plot the rescaled total traction force as a function of the amplification parameter, and observe a single master curve for all experiments, with varying $d$ and $N$.

The tremendous geometrical amplification of friction through the interleaving of the $2M$ sheets of each book bears a resemblance to two everyday examples. First, a sailor can moor a large ship simply by wrapping a rope around a cylinder known as a capstan~\cite{stuart1961capstan, attaway1999mechanics}. Note that its principle might be relevant to the interaction between DNA and a bacteriophage capsid~\cite{ghosal2012capstan}. The second example is a simple toy known as the Chinese finger trap, where a woven helical braid is loosely wrapped around a finger. The braid tightens and the finger is trapped as it is pulled. The trapping mechanism results from a simple conversion of the traction force to an orthogonal component, which enhances the load and thus the friction. This type of braid is applicable to sutures in surgery~\cite{smeak1990chinese}, and is also thought to play a role in adhesive proteins~\cite{le2010structural}. 

Crucial to the geometrical amplification of friction with interleaved books are the angles $\theta_n$ that the sheets make as they approach the contact region. It is through these angles that the traction forces result in loads perpendicular to the paper-paper interfaces, and thus in large self-created friction forces. This fact is easily verified: one can realise an interleaved-book system with $\theta_n=0$, by removing alternating sheets in two notepads. In such a case, the books can be easily pulled apart, consistent with the theory presented.

Finally, it is clear from the normalised data in Fig.~\ref{fig4} that the experiments deviate from the model at very small loads. This is the result of assuming a coefficient of kinetic friction that is independent of the load. While it is certainly possible to include numerically an \textit{ad-hoc} load-dependent coefficient of kinetic friction, this would be at the cost of simplicity and additional free parameters. In this Letter, we have instead opted to capture the essential physics needed to elucidate the interleaved-phonebook enigma, and we reveal the key dimensionless parameter $\alpha=2\mu\epsilon M^2/d$ of the problem.

\textbf{\\Acknowledgments} The authors thank the television channel France 5 and the 2P2L producing company, for the scientific program ``\textit{On n'est pas que des cobayes}''~\cite{cobayes_phonebook} which stimulated this study through the participation of F.R. and C.P., as well as the French ANR (WAFPI project: ANR-11-BS04-0030) and NSERC of Canada for funding.

\end{document}